# High resolution protein folding with a transferable potential.

Isaac A. Hubner[†], Eric J. Deeds[‡], and Eugene I. Shakhnovich[†*]

[†]Department of Chemistry and Chemical Biology

Harvard University

12 Oxford Street

Cambridge, MA 02138

[‡]Department of Molecular and Cellular Biology

Harvard University

7 Divinity Avenue

Cambridge, MA 02138

[*]corresponding author

tel: 617-495-4130

fax: 617-384-9228

email: eugene@belok.harvard.edu

**Abstract**

A generalized computational method for folding proteins with a fully transferable potential and geometrically realistic all-atom model is presented and tested on seven different helix bundle proteins. The protocol, which includes graph-theoretical analysis of the ensemble of resulting folded conformations, was systematically applied and consistently produced structure predictions of approximately 3Å without any knowledge of the native state. To measure and understand the significance of the results, extensive control simulations were conducted. Graph theoretic analysis provides a means for systematically identifying the native fold and provides physical insight, conceptually linking the results to modern theoretical views of protein folding. In addition to presenting a method for prediction of structure and folding mechanism, our model suggests that a accurate all-atom amino acid representation coupled with a physically reasonable atomic interaction potential (that does not require optimization to the test set) and hydrogen bonding are essential features for a realistic protein model.

**Introduction**

Protein folding is easy. Without effort, every living organism completes the process innumerable times. Unfortunately, modeling the process is notoriously difficult. Since Anfinsen's experiment[1], we have known that a protein's tertiary structure is defined by its primary sequence. However, the question of sequence-structure mapping remains unsolved. While researchers in the field have risen to the challenge and continue to make incremental progress[2], a complete solution remains among the great outstanding problems in computational biology. The problem has two aspects. First, given a protein's amino acid sequence, can one reliably predict its tertiary structure? Second, can one accurately understand and describe at a detailed atomic level the physical process by which a protein reaches it native conformation and the dynamics of the folded conformation?

Approaches to the protein folding problem fall into two major categories: bioinformatics methods attempt to model the structure of a protein primarily through homology to known structures and methods that rely on modeling the physical process by which the polymer chain attains its native conformation. While homology-based approaches have generally yielded more accurate structure predictions and are more readily applied to larger proteins[3], they do not provide physical insight into the folding or conformational dynamics of proteins. The "holy grail" of the protein folding community thus remains a computationally efficient model that both accurately predicts structure and provides physical insight into the folding and function of any protein given only its amino acid sequence.

Over the past decade, various models have been applied to protein folding and structure prediction. In an important study, the 36 residue villin headpiece fragment was folded to ~4-5Å from the native structure, demonstrating that the dream of *ab initio* protein folding is becoming a reality[4]. Other highly successful methods[5] combine sequence and structural homology with incremental physical model building for structure prediction. Many detailed physical studies use computationally intensive molecular dynamics (MD) simulations with complex potentials such as CHARMM and AMBER. Although they provide a measure of physical insight, these models have proven too computationally demanding to apply to any but the smallest proteins and, in such cases, usually produce results similar to simpler models. Additionally, the hundreds of extended simulations that would be necessary to create an ensemble picture of the folding process are beyond the reach of such models. This is especially important in context of the "new view" of protein folding as an ensemble process[6]. With the profound success of lattice models and Go-type energy potentials in studying and understanding protein folding[7-9], we know that simple models can effectively abstract many of the essential features of protein folding. We are thus encouraged that a similarly fundamental model, one that represents the basic physics of folding, accurately represents the protein structure (in real, not lattice space), and is not dependent on any *a priori* knowledge of the native fold may provide a solution to the folding problem.

The archetypal protein used in computational studies of folding is the B domain of *Staphylococcus* protein A [10]. Many papers have been devoted to computationally modeling this, and similar proteins, using everything from Go[11] potentials to empirical[3,12-14] and other potentials[15-19]. Simulations have utilized Monte Carlo (MC)[3,15], MD[13,14,18], discreet MD[11], and conformational space annealing[12]. Protein chain representations have varied from $C_\alpha$[11] and other reduced atom[12,16,18] to all-atom [3,13,14,19]models[3,13,14]. The structure prediction in these papers is generally in the 4Å range, which is similar to the best predictions of smaller proteins in CASP[2]. In most studies, the reported minimum RMSD conformation was not identifiable by energy[3,4,12-16,18-21]. Most current methods for protein folding rely on optimized Hamiltonians[17-19,21,22], similar in spirit to the methodology first introduced by Wolynes and coworkers[23]. Many such methods have > 100 adjustable parameters, raising questions about the transferability of the potential to proteins outside the training set.

Here, we present a novel model for high-resolution all-atom protein folding and demonstrate its efficacy on seven small helical proteins (three with albumin-binding topology, three with DNA binding topology[24], and a FF domain protein with one $3_{10}$ and three α helices). The model combines a realistic all-atom protein representation with a simple, fully transferable, contact potential[15] and hydrogen bonding function and is propagated *via* MC dynamics. An advantage of this method's computational efficiency is that it allows for hundreds of fully independent simulations, resulting in representative statistics and that allows one to test ensemble kinetics and thermodynamics. Importantly, the simulation requires no optimization and contains absolutely no knowledge of the protein's native structures and may be applied in a systematic prescribed manner to any amino acid sequence. Though hierarchical clustering has previously been utilized in protein structure prediction[25-27], we employ a different graph-theoretic approach that retains topological features of the relationships between members of a cluster and allows us to interpret the results in the context of landscape theories of folding, overcome noise in the potential, and identify high-resolution structure predictions from simulation without knowledge of the native state.

The resulting predictions, generally in the 3-4Å range, are significant by the criteria of the protein folding community[2] and on estimates based on studies of structural homology[28]. Additionally, we show that the results are meaningful when compared to a set of control simulations. Although conclusions regarding folding mechanism may be made from this model, we limit the present discussion to demonstrating the feasibility of folding and interpreting the graph theoretic analysis in terms of landscape theory. The successes of the model, which was not optimized or parameterized to any specific protein or training set, are derived from a realistic representation of the topological effects of folding: chain connectivity and sidechain packing combined with a simple, physically reasonable two-body potential that governs specific collapse and a generic hydrogen bonding potential that ensure secondary structure formation.

**Model and Methods**

**Protein representation and dynamics.** Simulations utilized structures from the protein data bank (PDB). For randomized sequence controls, all-atom models were built from the FASTA sequence using Swiss PDB Viewer. Non-hydrogen atoms are explicitly modeled as impenetrable hard spheres. The move set includes global and localized backbone moves and sidechain torsions (1 MC step is composed of 1 backbone move and 10 sidechain moves). Bond length and connectivity, as well as excluded volume are always maintained. The move set and MC simulation have been described in detail[29], and have been shown to behave ergodically and satisfy detailed balance[29,30]. Thought the thermodynamics and kinetics of the $\mu$-potential have not been calibrated in detail, cooperative two-state (single-exponential) folding and unfolding behavior is observed in all test proteins (data not shown).

**Form and derivation of the potential.** We have adapted and modified a transferable, knowledge-based pairwise contact potential from earlier work[15] that has also been used in a hybrid potential to introduce physically realistic interaction in simulations of SH3[31] of the form:

$$E_{AB} = \frac{-\mu N_{AB} + (1-\mu)\tilde{N}_{AB}}{\mu N_{AB} + (1-\mu)\tilde{N}_{AB}} \quad (1)$$

Where A and B are two interacting atoms, $N_{AB}$ and $\tilde{N}_{AB}$ are the number of AB pairs in contact and not in contact in the database, and $\mu$ is a parameter balancing attraction and repulsion. As is clear from the above equation, the potential becomes the Go potential at the limit where the number of atom types goes to the number of atoms in a single protein. The (non-optimized) value of $\mu$ (0.9979) was chosen such that the $<E_{AB}>$ = 0. To calculate $N_{AB}$ and $\tilde{N}_{AB}$, we use a database[32] of 103 proteins (Supplement 1) with <25% sequence homology that are longer than 50 and shorter than 200 residues and define a set of transferable atom types where backbone atoms are typed as peptide N, $C_\alpha$, carbonyl C, and O regardless of residue. Each sidechain atom has its own type, with the exception of those atoms related by symmetry (methyl group carbons in VAL, for example) yielding a total of 84 atom types (Supplement 2). Importantly, all proteins in our test set and their homologues are excluded from the database used to compute the potential. These two-body interactions are represented by a square well potential, where atoms A and B with hard-sphere radii $r$ separated by distance some $D$ are in contact if $0.75(r_A + r_B) < D < 1.8(r_A + r_B)$. The potential is available for download at http://www-

shakh.harvard.edu/~iahubner/pnas_supplement/supplement.html.

In addition to the above pairwise interaction potential, we consider a backbone hydrogen bonding ($E_{HB}$, Supplement 3) function to ensure proper secondary structure formation. The relative strength of hydrogen bonding and pairwise interaction is controlled by α, which balances the forces of polymer elongation and collapse.

$$E_{total} = \alpha E_{HB} + (1-\alpha) E_{AB} \quad (2)$$

In order to perform effective simulations, the relative energy scale between $E_{AB}$ and $E_{HB}$ must be set by α. When α is very high, the total energy is dominated by hydrogen bonding and extended helix conformations are formed. When α is too low, hydrophobic interactions, which are the strongest among $E_{AB}$ are overwhelmingly represented, leading to collapsed conformations with a well-packed hydrophobic core, but without secondary structure. At extreme (α ~ 0 or 1) values, protein behavior is sensitive to this parameter. However, it is possible to systematically identify an appropriate value of α by beginning with a high value and annealing until the majority of structures collapse to globular conformations (Supplement 4). The values of α that induce collapse are 0.92 for 1BDD, 1BA5, 1ENH, and 1GUU and 0.89 for 1GAB and 1GJS. Further tuning does not improve the results, and lowering α significantly beyond this point worsens the structures. As previously observed by other research groups[16], attempts at parameter optimization (in our case μ or α) did not improve the results.

**Simulation protocol.** First, the native PDB structure is unfolded for $10^6$ random (without energy, but maintaining excluded volume) MC steps at T=1000 to create a random, fully unfolded (extended and without correlations to native ϕ and ψ angles) starting conformation. Each folding simulation is initiated from an independent random conformation and propagated at T=1.75 ($\sim T_f$) for $10^8$ MC steps. Next, the minimum energy conformation from the folding simulation is annealed from T=1.75 to 0 for $5 \cdot 10^7$ MC steps to improve sidechain packing. The minimum energy conformation from this refinement simulation is then collected as the structure prediction. The above protocol was repeated 400 times for each protein to create an ensemble of predictions.

**Test proteins.** Seven independently folding domains of short length were selected to test the folding model. The albumin binding ("up-down-up" three helix bundle) topology, which includes: *Staphylococcus Aureus* protein A, immunoglobulin-binding B domain (1BDD), *E. coli* Albumin-binding domain surface protein (1GAB), and *Streptococcus* Immunoglobulin G binding protein G (1GJS). The DNA binding ("helix-turn-helix" or homeodomain-type) topology, including: DNA binding domain of human telomeric protein HTRF1 (1BA5), an engrailed homeodomain-DNA complex (1ENH), and c-Myb R1 proto-oncogene (1GUU). An FF domain protein (1UZC), with a more complex four-helix (one of which is $3_{10}$) topology was also included. 1ENH and 1GUU were compared to X-ray structures, whereas all others were compared to the NMR structures. Because all proteins had long, disordered regions at the termini, we calculate RMSD for the helical and turns regions, corresponding to F6-A60 in 1BDD, N9-A53 in 1GAB, D16-A62 in 1GJS, L7-L53 in 1BA5, F8-I56 in 1ENH, T44-L86 in 1GUU, and K14-T69 in 1UZC. This choice of test set allows for the comparison of different folds and multiple sequences of the same topology.

**Graph-theoretical analysis.** The RMS deviation in $C_\alpha$ coordinates is computed for all pairs of the lowest energy structures obtained from 400 independent simulations of each protein. A graph is then created from these comparisons by considering each minimum energy structure as a node and connecting any two nodes that exhibit an RMSD less than a particular cutoff ($r$). The clusters in this graph are defined as any set of nodes where a path exists in the graph between any two members of that set. These disjoint clusters are obtained using a standard depth-first search algorithm. At any given value of $r$, the Giant Component (GC) of the graph is defined as the largest disjoint cluster. As has been observed in many systems, this GC undergoes a transition as a function of $r$ (Supplement 6). We analyze the GC for each protein at the mid-point of the transition, the cutoff $r$ at which half of the total structures are contained within the largest cluster.

## Results and Discussion

**Separating native folds from misfolds.** From examining the 400 independent trajectories for each protein, it is clear that the native state is well sampled. Most minimum energy conformations fall in the 2-6Å range. Examining all minimum energy structures (Table 1), it is clear that along with native folds, there are a number of low energy decoys. Due to the approximate nature of our energy function, this is not surprising. Previous computational studies also resulted in the minimum of various energy functions not corresponding exactly to the native state[3,4,12-16,20,21]. Many of the decoys we observe involve undocked helices or poorly formed secondary structure. These misfolds usually exhibit higher energy than near-native structures and are thus easily identifiable. There is a second class of misfolds that are protein-like and are broadly describable as "mirror-images" of the native structure. These misfolds represent a more difficult case since they have energies comparable to native-like structures despite high (8-10Å) RMSD from the native state. Other researchers have also noted the presence of mirror misfolds, as a three helix bundle exhibits twofold topologic degeneracy[3,12,16].

Given that our energy function cannot *a priori* distinguish these low-energy decoys from native conformations we must rely on other objective analyses to identify native conformations. To accomplish this we employ a graph theoretic clustering procedure. The largest cluster in this graph, the GC, undergoes a sharp transition as $r$, the structural cutoff employed to construct the graph, becomes more stringent. At the midpoint of the transition many of the decoy structures are excluded from the GC as evidenced by the decrease in average RMSD in the GC compared to the entire ensemble of structures at the mid-point of the transition in the GC (Table 1). In most cases the mirror misfold structures are excluded from the GC at this point in the transition. A representative graph (Supplement 7) for 1BDD at the midpoint in the transition is shown in Figure 2. As evidenced by these graphs, in the predominance of cases no set of decoys obtained from these simulations forms a cluster that is as large and coherent as the native-like structures in the GC. This indicates that the predominant structural class sampled by our simulations and identified *via* clustering is the native basin.

The two exceptions to the above observation are 1GUU and 1ENH. In the case of 1GUU mirror misfolds remain in the GC at the midpoint of the transition. Visual inspection of the 1GUU graph at the transition (Supplement 7) clearly reveals that the GC in this graph consists of two distinct, dense clusters that are connected to one another by only a few edges. At a lower cutoff these two clusters break into a near-native cluster and a mirror misfold cluster (Table 3). In the case of 1ENH the misfolds form a coherent cluster separate from, but of almost the same size as, the GC (Supplement 7 and Table 3). In both of these cases our method cannot identify which represents the native cluster. Graph theoretic analysis allows us to identify this problem when it does occur, however, providing an objective measure of the degeneracy in our sampling of structural space.

**Identifying the native fold.** Clustering the results of independent folding simulations improves the quality of the prediction by enriching the representation of the well-sampled native state and excluding the disparate misfolds. However, the size of the GC may be quite large (200 of 400 configurations at the half point of the transition). This raises the question of how the best models from the GC be reliably chosen using objective, quantifiable criteria. While ranking predictions by energy provides reasonable results, it clearly fails in such cases as 1GAB (Table 2). The success of clustering in eliminating misfolds suggests that some topological features of the graphs may serve to identify the native state. We hypothesize that the most native conformations, if properly sampled, should be the most connected within the GC since a higher population of similar structures results in more connections between those conformations. We find that clustered conformations exhibit a general relationship between the number of neighbors a node exhibits (called $k$, the degree of the node) and the RMSD from the native state (supplement 8). Although conformations with low RMSD from the native may exhibit a low degree, most nodes of high degree are among the most native conformations observed in our simulations. When the 3 highest degree nodes from the GC are chosen they unilaterally include some of the highest quality structures from the entire simulation.

This approach is only possible because the graph-theoretic approach we employ preserves topological information about the relationship between nodes in a cluster.

The superposition of the top *k* prediction and native state for each protein is presented in Figure 1. These predictions are obtained from a generalized, fully transferable potential, and the graph-theoretic approach represents a completely consistent, objective analysis that requires no knowledge of the native state. The resulting predictions, three around 3Å RMSD, three at 4Å, and a 5Å "fold prediction" in the worst case, demonstrate the effectiveness of our potential and the utility of the clustering method. As discussed in the preceding section, the 1GUU GC contains native and misfolds at the half point and splits into two clusters at lower cutoffs. Applying the same highest *k* criteria to these clusters provides a very good (3Å) structure prediction in one of the two cases. In every test case the GC at half transition contains the native fold. In every test, the native fold is always among the highest 3 *k* conformations. Graph theoretic analysis thus provides a means for enriching the native fold *via* clustering, choosing a high-quality prediction using *k*, and gauging the reliability of that prediction by identifying the misfold problem when it occurs.

**Controls and the meaning of the results.** From the application of our protocol to seven real proteins, it is clear that sequences designed by evolution to fold do so in our model. However, there are several potential concerns that should be addressed through control simulations. First, we show that the μ-potential contains useful and meaningful information by demonstrating that it does not behave like a random potential. Second, we confirm that the μ-potential predictions are meaningful in that the model is not constructed to generically fold every sequence into a helix bundle topology. Last, we verify that the predicted topologies result from the interplay between the hydrogen bonding and pair contact potential.

In all control simulations (Table 4), 1BDD is selected to represent albumin-binding domains and 1ENH to represent DNA-binding domains. The protocol for 200 folding simulations and clustering analysis for each control, are identical to those for folding the test proteins. The µ-potential has interaction energies that range from -1 to 1 and have an average of 0. For comparison, we produce a random potential with the same statistical properties and apply it, along with the same H-bond potential, to the folding of the 1BDD and 1ENH sequences. In 1BDD, this results in an average of 16.72Å RMSD from the native state and for 1ENH the average RMSD of the resulting structures is 16.40Å. In both cases, the structure nearest to native is ~11Å and, upon clustering, the GC does not improve any of the predictions. None of the folding runs produced a three-helix bundle topology even remotely similar to the native conformation. Clearly, the µ-potential is non-random and contains information that discriminates the native protein structure.

In order to demonstrate that the model is not contrived to produce helix bundle proteins, we show that if the protocol is applied to some arbitrary random protein-like sequence it does not produce helix bundle conformation. The FASTA sequences for 1BDD and 1ENH were each subjected to 6000 randomizing permutations. When the resulting sequence was checked using BLAST against the non-redundant NCBI database of protein sequences[33], no known sequence homologues were found. A new amino acid chain of the same amino acid composition and length as two different real proteins, but different sequence was then constructed using Swiss PDB View. From the results it is clear that the 200 control runs result in structures that are much worse than the prediction runs and that no improvement comes from clustering. Within the GC, no helix bundle topology resembling the native conformation of either protein is identified by energy or clustering (Table 4). Though no conformations resembling the WT are produced, the GC and "prediction" represent a cohesive set of similar structures. Clearly, this model is not contrived to turn any sequence into a helix bundle. Based on these data, a random collapsed false prediction should exhibit a RMSD of ~10Å, highlighting the significance of the ~3Å predictions.

Simulations run with only the μ-potential, result in collapsed structures with a compact hydrophobic core, but without secondary structure, that are ~10Å RMSD from the native state. Moreover, clustering analysis fails to identify a dominant GC, even at high $r$ cutoffs. When the conformations do cluster, they form a large number of small clusters, indicating great structural diversity. If simulations are run with only hydrogen bonding, then the amino acid chain becomes a single extended helix, with a 19.43 and 21.30Å RMSD for 1ENH and 1BDD, respectively. Neither $E_{AB}$ nor $E_{HB}$ can alone identify native conformations. While an atomic interaction terms is necessary for collapse, secondary structure formation functions to limit the conformational space available to an amino acid chain, reducing the conformational space necessary to search for the global minimum and contributing to the presence of a cohesive, well defined ensemble conformations in a protein's folded state. It was recently suggested[34,35] that compaction, chain geometry, and excluded volume ensure protein-like conformations. The above simulations indicate that these factors alone are not able to produce protein-like conformations. We find that necessary conditions for a protein model to achieve a realistic tertiary structure include a geometrically and spatially realistic sidechain and backbone representation, an accurate representation of hydrogen bonding, and a potential that represents specific hydrophobic and other atom-atom interaction in a physically appropriate manner.

From the clustering of each control (Table 4) it is clear that the resulting "predictions" are non-random, but do not resemble helix bundles. We observe primarily compact conformations that may have helices and turns, with the obvious exception of using only the hydrogen bonding potential, which produced a single extended helix. It is not surprising to observe a small number of collapsed conformations resembling helix bundles in the RS controls, some as close as 4Å from the native. However these conformations cannot be identified by energy or any graph-theoretical criteria introduced in this work. Several studies have based claims of successful folding, at least in part, on the presence of low RMSD conformations[13,36] in a subset of trajectories. However, from these controls we see that it is possible to sample helix bundle conformations by randomly collapsing a random sequence heteropolymer chain with the caveat that the resulting native-like structure exists only in a tiny fractions of runs and cannot be identified by clustering or energy. For a protein folding simulation, it is not enough to identify "native" conformations by low RMSD; only identification of low RMSD conformations by quantitative criteria independent of knowledge of the native state constitutes a successful description of a physical sequence-dependent folding event. The structure predictions identified by energy and graph properties are highly nontrivial, especially when compared to the ~10Å control results. In each case of an evolved protein sequence, the model produces, and clustering is able to discriminate a single well-populated minimum energy conformational class, demonstrating the utility of clustering independent of the potential chosen. The potential contains meaningful and useful information, and is not constructed to turn any sequence into a protein-like state or helix bundle. The protein model effectively represents protein behavior for native sequences and structures by modeling amino acid conformational space, hydrogen bonding, and atom-atom interactions.

**Conclusions.**

These results represent significant progress and promise for understanding protein folding and structure prediction in *ab initio* high-resolution simulations. Systematically applying a quantitative method that uses an energy function along with clustering to identify the prediction consistently identifies structures in the 3Å RMSD range for most proteins. Clustering works in concert with the ensemble of energy-based predictions to reliably eliminate decoys and identify the free energy minimum structures that closely resemble the native state. Graph-theoretic analysis provides an indication of the quality of the results, identifies the most native folds and misfolds, and provides a conceptually useful link to interpretation of the results in the context of physical theories of folding. Future extensions of this work include improving the discriminatory ability of the potential and including solvent effects as well as the extension to larger, diverse proteins, structural refinement of fold predictions from homology modeling, and predicting protein folding mechanisms.

Figure 2 shows the graph for all structures at the transition in the giant component for 1BDD folding simulations and randomized sequence control. It is clear from this figure and it as has been shown analytically[37,38], that random and designed heteropolymers exhibit different behaviors dictated by their energy landscapes. Whereas the designed (protein-like) polymer largely populates a distinct and deep minimum, the random polymer inhabits multiple, energetically similar, but structurally unrelated states. This view from heteroplymer theory explains why clustering of conformations improves predictive power of the method: native-like states feature multiple interactions that work in concert to provide minimum energy structure in the native state, consistent with the "principle of minimal frustration"[39]. Since our potential is noisy and approximate, the energy landscape has features of both design and random heteropolymers. Apparently the native basin of attraction is ''broad'' containing many structurally related conformations, in contrast to spurious minima, characteristic of random heteropolymer that may be deep but contain only few structures. A graph is a topological entity and can serve to analyze and conceptualize the multi-dimensional protein folding energy landscape[25,40] without reliance on spatial coordinates and, as such, has potential for representing energy landscapes in a limited set of order parameters. These data show that a high *k* in a graph corresponds to a high density of states, or energy minimum, on the protein folding energy landscape. In addition to understanding energy landscape topology at the minimum (structure prediction), we anticipate many applications for this model in understanding global landscape topology (folding).

We have demonstrated that fully atomistic simulations, using a single protein representation from beginning to end, have the ability to fold multiple proteins to their native states with a single transferable potential that is not trained or optimized on the test proteins or decoy sets and relies on absolutely no information of the native structure. Our results are comparable with studies of test-set optimized potentials and go beyond published studies of non-optimized potentials (Table 5). Controls show that simply collapsing an amino acid chain with an attractive pair potential is insufficient to fold proteins; the problem of secondary structure must also be addressed. Likewise, we have demonstrated that our potential contains enough specific information to identify the native structures of proteins, without such bias that it introduces false positive predictions – it does not turn any sequence into a helix bundle protein. The model works because it accurately captures protein geometry and sidechain packing, hydrogen bonding and secondary structure formation, and presents a physically reasonable pairwise potential for compaction. It is both encouraging and intellectually satisfying that simple physical models reliably represent many of the aspects of protein folding and that graph theoretic analysis conceptually links the results of folding simulations to energy landscape theory.

**Acknowledgement.**

We thank C. Brian Roland for simulating discussion and careful reading of the manuscript. IAH and EJD are supported by the Howard Hughes Medical Institute. This work is supported by NIH grant GM52126.

**References.**

Table 1. The Giant Component presents a significant enrichment of the data, eliminating misfolds and reducing the average RMSD while retaining the best predictions. The "All" column describes the average and range of RMSD values for the $E_{min}$ conformation from each of 400 independent simulations; GC values are calculated at the transition midpoint.

| Protein | All | | Giant Component | |
|---|---|---|---|---|
| | <RMSD> | Range | <RMSD> | Range |
| 1BDD | 9.21 | 3.00-19.71 | 6.01 | 3.00-10.16 |
| 1GAB | 7.26 | 2.56-11.99 | 6.35 | 2.56-9.86 |
| 1GJS | 7.51 | 2.43-16.63 | 5.25 | 2.43-8.73 |
| 1BA5 | 8.33 | 3.79-16.76 | 6.31 | 3.79-8.94 |
| 1ENH | 9.19 | 2.44-18.40 | 5.42 | 2.44-9.41 |
| 1GUU | 8.07 | 3.28-13.95 | 7.14 | 3.28-10.06 |
| 1UZC | 8.75 | 3.42-12.15 | 7.72 | 3.42-11.44 |

Table 2. RMSD of structure predictions from the Giant Component by $E$ and $k$.

|  | 3 lowest $E$ | 3 highest $k$ |
|---|---|---|
| 1BDD | 4.00, 4.25, 6.82 | 4.77, 5.55, 6.27 |
| 1GAB | 7.47, 7.71, 8.20 | 3.07, 3.10, 3.41 |
| 1GJS | 4.51, 5.07, 5.36 | 3.76, 3.97, 4.65 |
| 1BA5 | 5.82, 5.92, 6.59 | 5.63, 5.72, 5.85 |
| 1ENH | 2.96, 4.63, 5.72 | 2.44, 2.96, 3.01 |
| 1GUU | 7.73, 8.62, 9.17 | 4.05, 8.25, 8.99 |
| 1UZC | 5.05, 6.10, 7.09 | 4.25, 9.22, 9.40 |

Table 3. Comparison of native and mirror misfold clusters. For 1ENH, 1 is the GC at half transition and 2 is another large cluster. GUU clusters 1 and 2 presented here are at a cutoff of 2.6Å (after the midpoint), where the two folds separate.

|  | Giant Component | | Prediction | |
| --- | --- | --- | --- | --- |
|  | <RMSD> | Range | 3 lowest $E$ | 3 highest $k$ |
| ENH 1 | 5.42 | 2.44-9.41 | 2.96, 4.63, 5.72 | 2.44, 2.96, 3.01 |
| ENH 2 | 9.85 | 9.03-10.73 | 9.03, 9.51, 9.86 | 9.30, 9.64, 9.86 |
| GUU 1 | 4.59 | 3.28-6.37 | 4.19, 4.38, 4.43 | 3.40, 3.93, 4.41 |
| GUU 2 | 8.82 | 7.36-9.81 | 7.73, 8.62, 9.17 | 8.97, 8.99, 9.29 |

Table 4. $C_\alpha$ RMSD results from controls of 1ENH and 1BDD. The randomized sequence (RS), randomized potential (RP), and $E_{AB}$ potential only (PO) yield poor results that are not improved by graph-theoretical analysis.

|  |  | All 200 |  | Giant Component |  | "Prediction" |  |
|---|---|---|---|---|---|---|---|
|  |  | <RMSD> | Range | <RMSD> | Range | 3 lowest $E$ | 3 highest $k$ |
| 1ENH | RS | 10.70 | 4.60-16.10 | 10.25 | 9.18-11.47 | 9.69, 10.18, 10.37 | 9.95, 10.26, 10.43 |
|  | RP | 16.40 | 11.2-19.25 | 16.30 | 14.52-18.33 | 14.51, 15.73, 17.86 | 14.93, 15.20, 16.05 |
|  | PO | 9.67 | 5.66-13.07 | 10.24 | 8.78-11.03 | 9.55, 10.63, 11.00 | 10.12, 10.22, 10.52 |
| 1BDD | RS | 12.72 | 4.77-22.09 | 9.62 | 6.13-12.05 | 11.07, 11.42, 12.05 | 10.18, 10.42, 11.42 |
|  | RP | 16.72 | 10.65-22.10 | 17.98 | 15.98-20.46 | 16.26, 18.63, 19.50 | 18.26, 18.63, 18.85 |
|  | PO | 10.70 | 7.15-13.68 | 10.52 | 8.60-13.07 | 9.38, 10.94, 11.02 | 9.98, 10.16, 10.47 |

Table 5. Comparison of representative, contemporary models for protein folding. $C_\alpha$ RMSD to native PDB (rounded to nearest whole number) is for best objective predictions, <u>not</u> the lowest value observed in simulation. Methods with "optimized" potentials were either trained on test proteins' native state or decoys of the test proteins or other databases.

| reference | Hubner *et. al.* | 17 | 19 | 18 | 14 | 22 | 13 | 36 | 16 | 21 |
|---|---|---|---|---|---|---|---|---|---|---|
| All atom? | Y | N | Y | N | Y | N | Y | Y | N | N |
| Length of protein(s) | 43-56 | 28-75 | 20-60 | 47-76 | 46 | 36-147 | 36-46 | 36 | 46 | 46 |
| Number of proteins | 7 | 7 | 4 | 6 | 1 | 11 | 2 | 1 | 1 | 1 |
| RMSD (Å) prediction(s) | 2-5 | 2-5 | 2-4 | 4-11 | * | ** | *** | **** | 9 | 3 |
| Optimized? | N | Y | Y | Y | N | Y | N | N | N | Y |

\* RMSD of $E_{min}$ structure was not given, but 3-4Å conformations were observed. However, some replicas were initiated from <2Å conformations.
\*\* Required input of experimentally determined secondary structures and optimization on individual proteins to attain 2-4Å
\*\*\* RMSD of $E_{min}$ structures was not given, but the 3-6Å range was "populated"
\*\*\*\* no RMSD (only minimum dRMS to a "relaxed" native state ensemble) reported

Figure 1. Top *k* predictions from Table 2, superimposed on the native conformation and colored from N (blue) to C (red) termini.

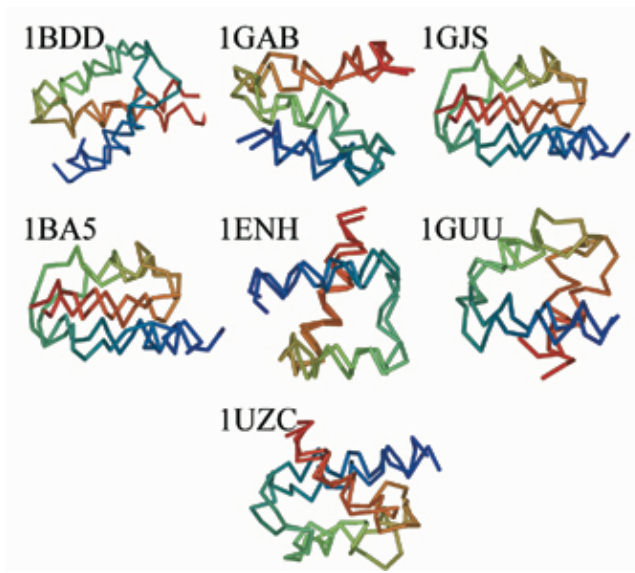

Figure 2. The relationship between graph-theoretic analysis and landscape theory. Proteins (designed heteropolymers) exhibit a deep, pronounced minimum (large, dense native cluster), but the landscape is rugged with low energy traps (other small and disjoint clusters). The randomized sequence behaves as a random heteropolymer, characterized by multiple energy minima and lacking a single, prominent minimum (native conformation). Graphs are calculated with results from simulations; landscape cartoons are only illustrative.

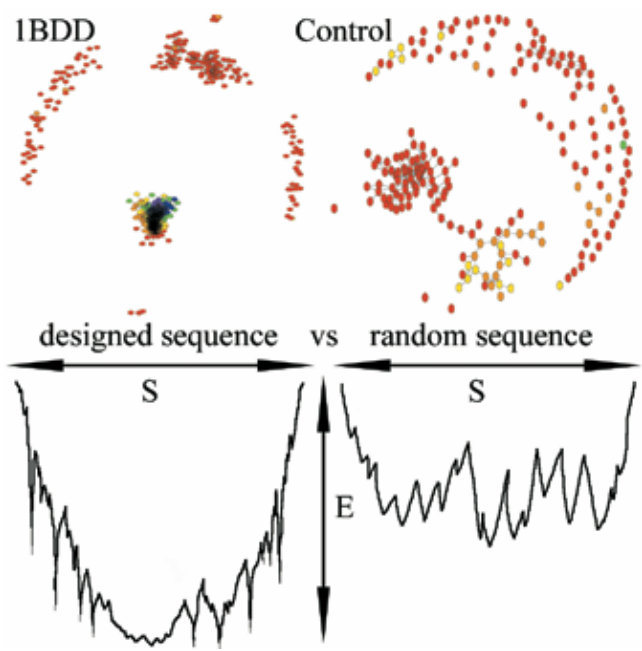

Supplement 1.  103 protein database.

119l, 1aya, 1ccr, 1f3g, 1hbi, 1ilr, 1onc, 1r69, 2alp, 2mta, 4icb, 135l, 1bab, 1cdl, 1fdd, 1hbq, 1ith, 1ovo, 1rcb, 2asr, 2rn2, 5p21, 1a45, 1bbh, 1cob, 1fha, 1hcr, 1lba, 1pal, 1rcf, 2ccy, 2rsl, 8atc, 1aap, 1bbp, 1cse, 1fkb, 1hfc, 1lmb, 1paz, 1rpg, 2cpl, 2sic, 9wga, 1aba, 1bet, 1cyo, 1flp, 1hjr, 1lpe, 1pk4, 1sha, 2end, 2sn3, 1acf, 1bov, 1dyn, 1fna, 1hlb, 1mba, 1plc, 1slc, 2fox, 2trx, 1acx, 1brn, 1eco, 1frd, 1hsb, 1mbd, 1plf, 1stf, 2fx2, 3chy, 1adl, 1brs, 1fus, 1hst, 1mjc, 1pnt, 1ubq, 2gmf, 3dfr, 1aiz, 1c2r, 1esl, 1fxd, 1hyp, 1mol, 1poh, 256b, 2hbg, 4dfr, 1ash, 1cad, 1etb, 1gmp, 1ida, 1ndc, 1psp, 2aak, 2msb, 4i1b

Supplement 2. 84 atom types.

| atom | residue | type | atom | residue | type |
|---|---|---|---|---|---|
| CB | ALA | 0 | CB | MET | 43 |
| CB | ARG | 1 | CG | MET | 44 |
| CG | ARG | 2 | SD | MET | 45 |
| CD | ARG | 3 | CE | MET | 46 |
| NE | ARG | 4 | CB | PHE | 47 |
| CZ | ARG | 5 | CG | PHE | 48 |
| NH1 | ARG | 6 | CD1 | PHE | 49 |
| NH2 | ARG | 6 | CD2 | PHE | 49 |
| CB | ASN | 7 | CE1 | PHE | 50 |
| CG | ASN | 8 | CE2 | PHE | 50 |
| OD1 | ASN | 9 | CZ | PHE | 51 |
| ND2 | ASN | 10 | CB | PRO | 52 |
| CB | ASP | 11 | CG | PRO | 53 |
| CG | ASP | 12 | CD | PRO | 54 |
| OD1 | ASP | 13 | CB | SER | 55 |
| OD2 | ASP | 13 | OG | SER | 56 |
| CB | CYS | 14 | CB | THR | 57 |
| SG | CYS | 15 | OG1 | THR | 58 |
| CB | GLN | 16 | CG2 | THR | 59 |
| CG | GLN | 17 | CB | TRP | 60 |
| CD | GLN | 18 | CG | TRP | 61 |
| OE1 | GLN | 19 | CD1 | TRP | 62 |
| NE2 | GLN | 20 | CD2 | TRP | 63 |
| CB | GLU | 21 | NE1 | TRP | 64 |
| CG | GLU | 22 | CE2 | TRP | 65 |
| CD | GLU | 23 | CE3 | TRP | 66 |
| OE1 | GLU | 24 | CZ2 | TRP | 67 |
| OE2 | GLU | 24 | CZ3 | TRP | 68 |
| CB | HIS | 25 | CH2 | TRP | 69 |
| CG | HIS | 26 | CB | TYR | 70 |
| ND1 | HIS | 27 | CG | TYR | 71 |
| CD2 | HIS | 28 | CD1 | TYR | 72 |
| CE1 | HIS | 29 | CD2 | TYR | 72 |
| NE2 | HIS | 30 | CE1 | TYR | 73 |
| CB | ILE | 31 | CE2 | TYR | 73 |
| CG1 | ILE | 32 | CZ | TYR | 74 |
| CG2 | ILE | 33 | OH | TYR | 75 |
| CD1 | ILE | 34 | CB | VAL | 76 |
| CB | LEU | 35 | CG1 | VAL | 77 |
| CG | LEU | 36 | CG2 | VAL | 77 |
| CD1 | LEU | 37 | CA | GLY | 78 |
| CD2 | LEU | 37 | N | XXX | 79 |

| CB | LYS | 38 | CA  | XXX | 80 |
|----|-----|----|-----|-----|----|
| CG | LYS | 39 | C   | XXX | 81 |
| CD | LYS | 40 | O   | XXX | 82 |
| CE | LYS | 41 | OXT | XXX | 83 |
| NZ | LYS | 42 | OCT | XXX | 83 |

Supplement 3. Schematic representation of $E_{HB}$ interaction. A hydrogen bond is counted when the four atom pairs are within a square well, eliminating the need for angle calculations and increasing computational efficiency. The indicated distances are as follows: d1 is between the donor nitrogen and acceptor oxygen, d2 is between the donor nitrogen and acceptor carbonyl carbon, d3 is between the donor hydrogen and acceptor oxygen, and d4 v is between the donor hydrogen and acceptor carbonyl carbon.

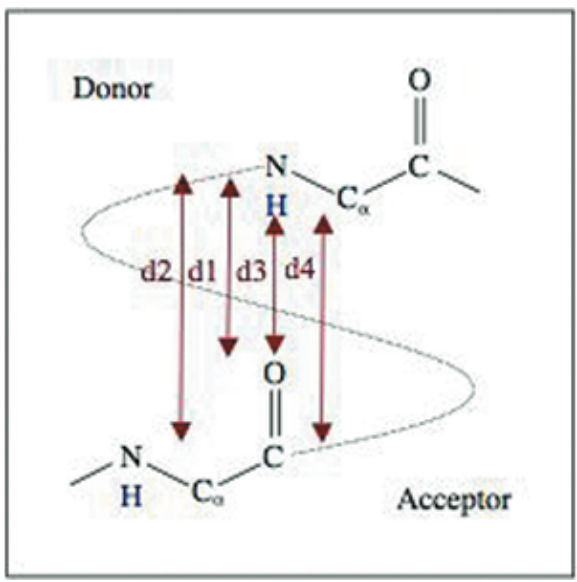

Supplement 4. A plot of *Rg vs* α for 1GJS reveals that systematically lowering α induces collapse (as measured by *Rg*, which requires no knowledge of the native state) and that lowering α beyond this point does not alter compaction. We empirically observe that lowering α significantly below this point reduces secondary structure, as hydrogen bonding will contribute less to the total energy. Therefore, α should be selected at the point at which the average over all structures reflects collapse, but not any lower (0.89 in the case below, see arrow). Since protein folding behavior is not very sensitive to small changes in α (the fluctuation between runs at a given α are much larger than the differences between runs with small (<0.01) differences in α), parameter optimization is neither useful nor necessary beyond finding the point at which average collapse occurs.

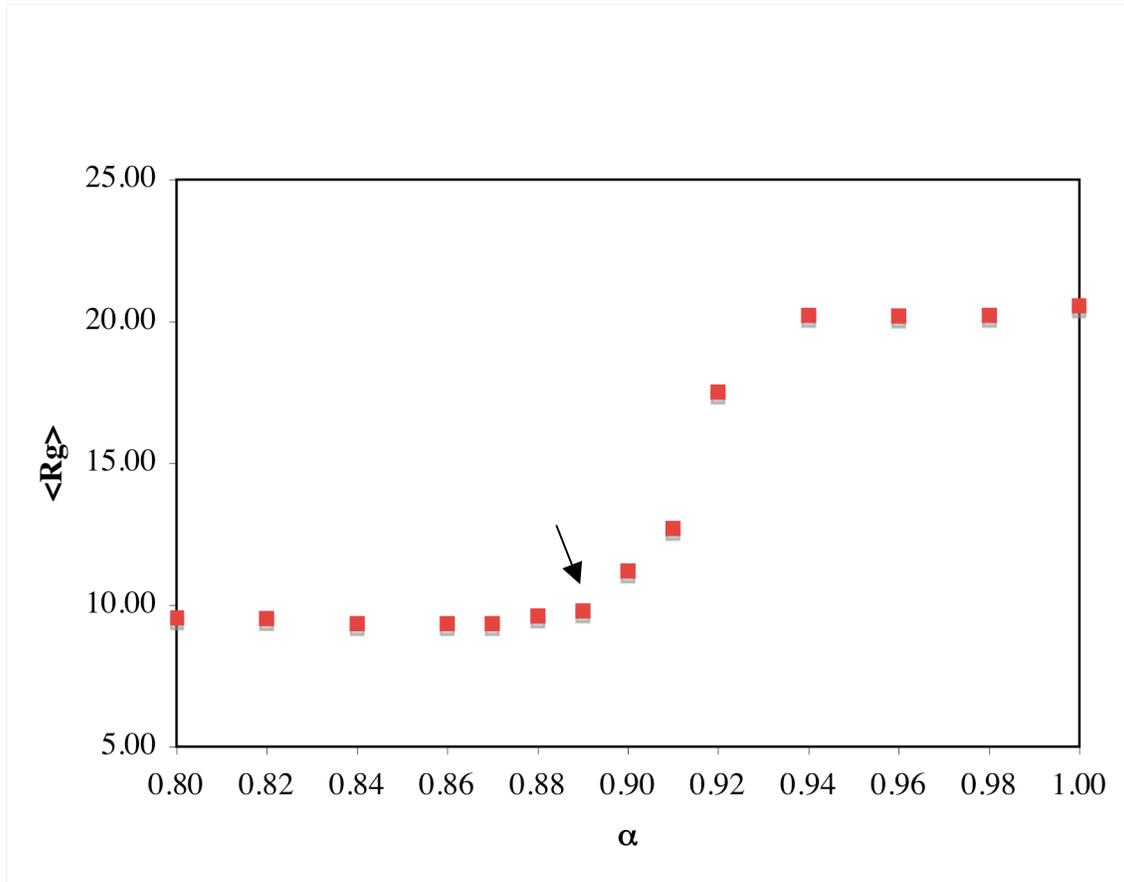

Supplement 5. Percent sequence identity between test proteins.

|      | 1BDD | 1GAB | 1GJS | 1BA5 | 1ENH | 1GUU | 1UZC |
|------|------|------|------|------|------|------|------|
| 1BDD | 100  | 5    | 18   | 5    | 3    | 2    | 15   |
| 1GAB |      | 100  | 50   | 5    | 7    | 6    | 5    |
| 1GJS |      |      | 100  | 9    | 7    | 12   | 50   |
| 1BA5 |      |      |      | 100  | 3    | 30   | 5    |
| 1ENH |      |      |      |      | 100  | 10   | 7    |
| 1GUU |      |      |      |      |      | 100  | 5    |
| 1UZC |      |      |      |      |      |      | 100  |

Supplement 6. Transition in the giant component for six test protein graphs.

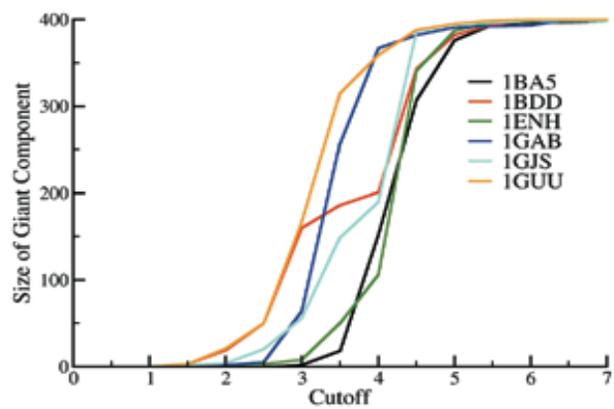

Supplement 7. Graphs for 1GAB, 1GJS, 1BA5, 1ENH, and 1GUU predictions. Nodes are colored to represent the RMSD from the native state with RMSD < 3.5 purple, <4.0 blue, <4.5 green, <5.5 yellow, <6.5 orange, >6.5 red.

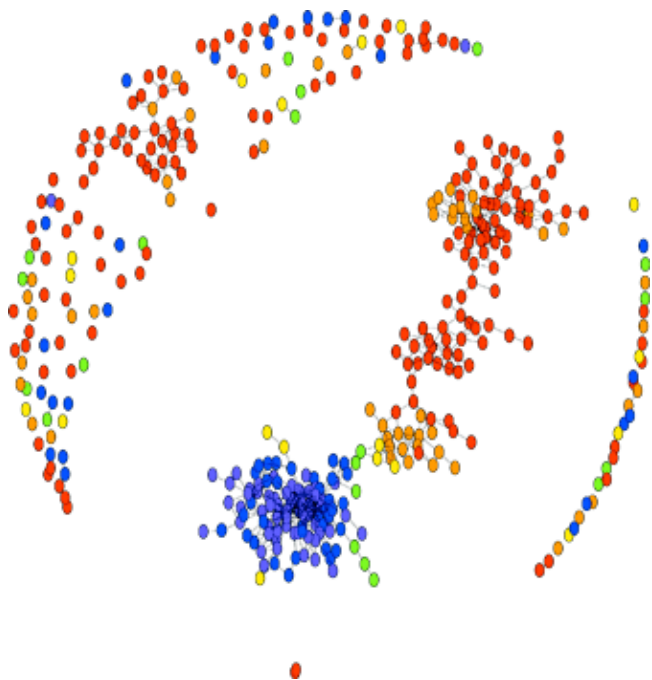

1GAB

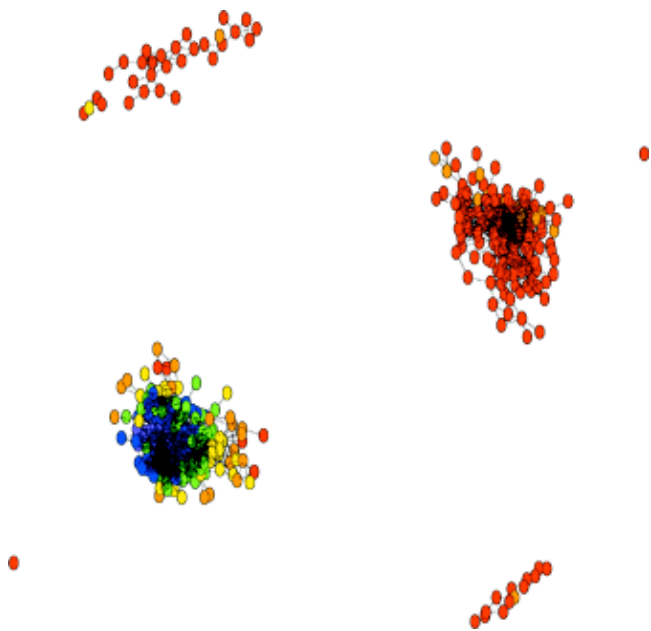

1GJS

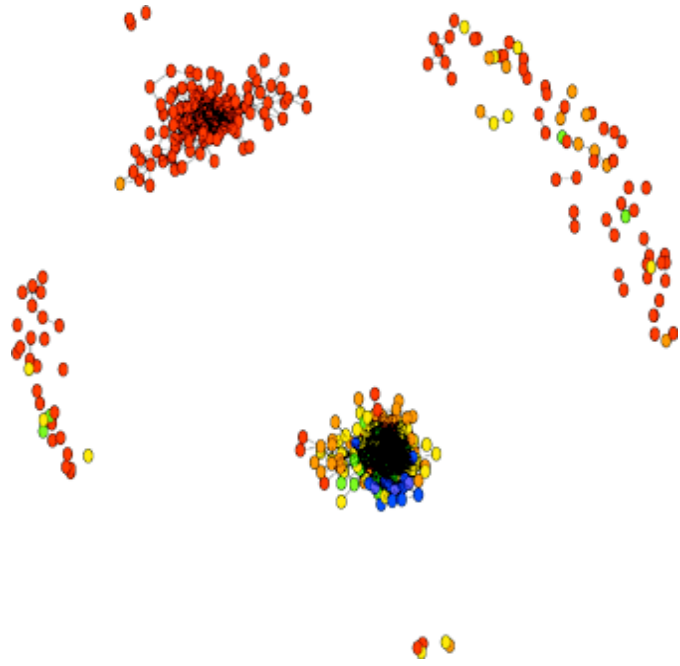

1BA5

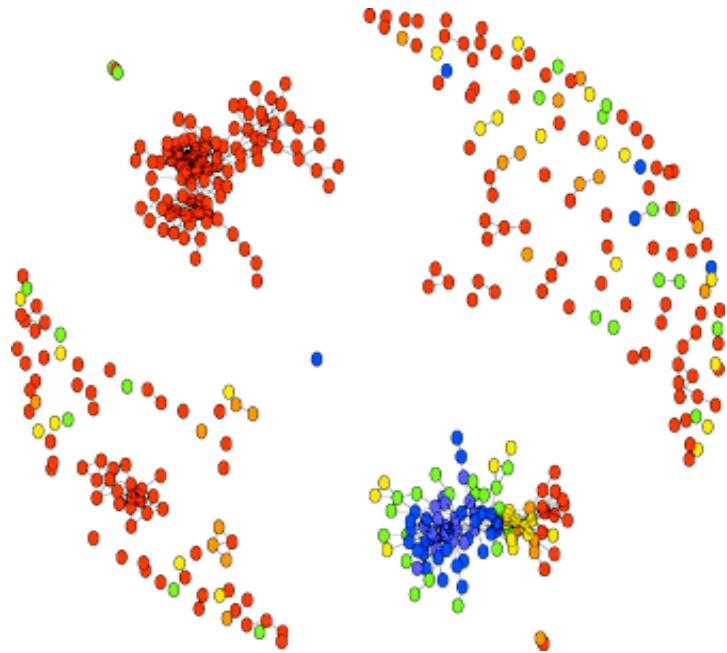

1ENH

1GUU

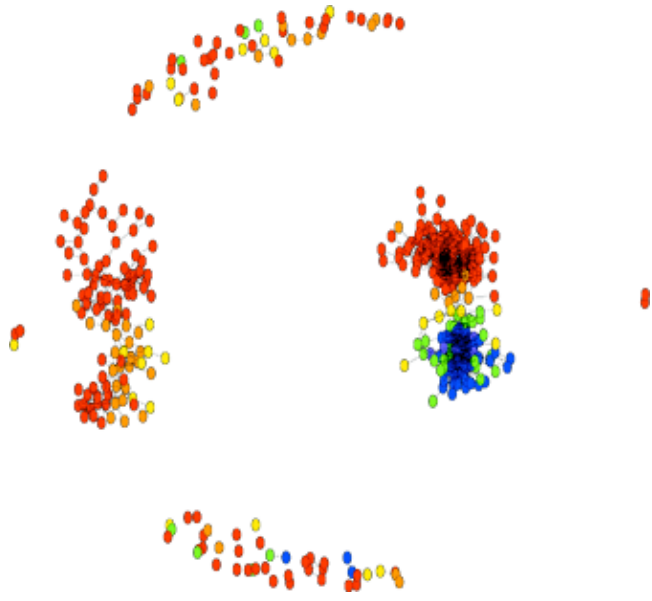

Supplement 8. *k* vs. RMSD within the GC of the six test proteins. Although the correlation is not perfect, k is a reliable predictor of the lowest RMSD conformations within a cluster.

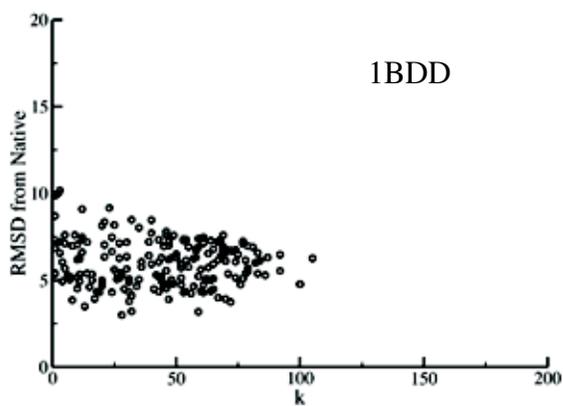

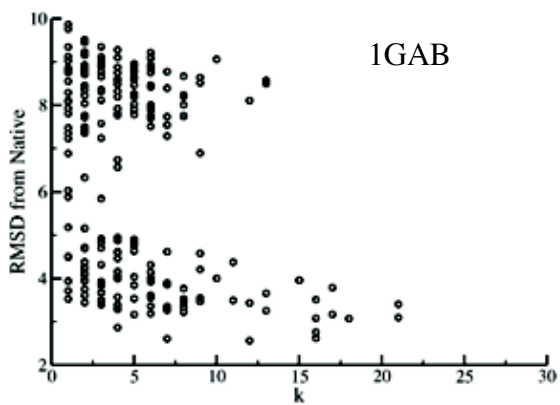

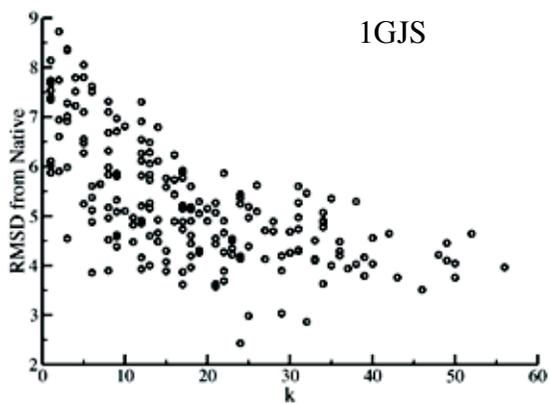

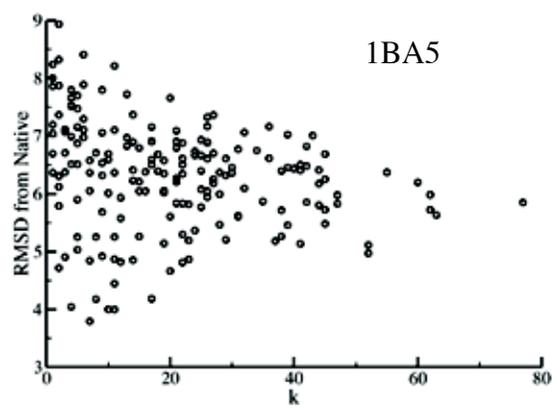

1BA5

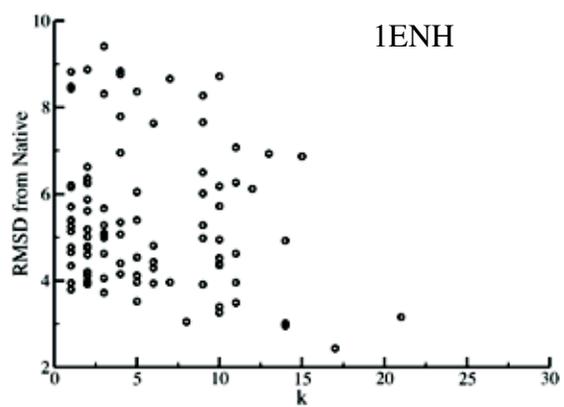

1ENH

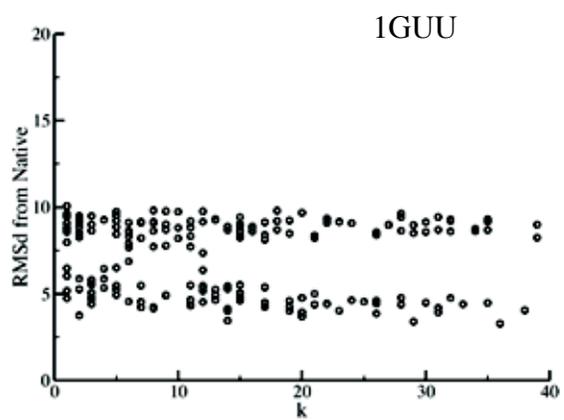

1GUU

Supplement 9. 1ENH Randomized control graph. Color scheme is the same as Supplement 6.

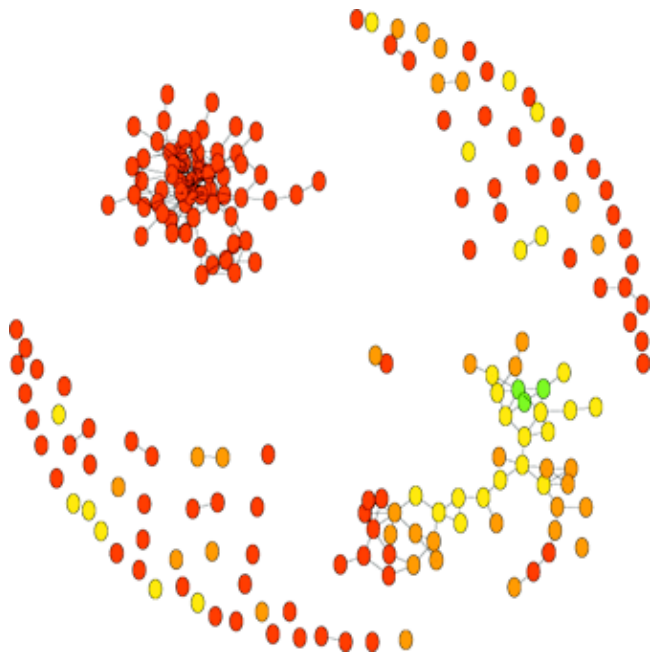